\documentclass[aps,preprintnumbers,twocolumn]{revtex4}
\usepackage{amsfonts,amssymb,amsmath}            
\usepackage[pdftex]{graphics,graphicx,epsfig}            
\usepackage{amsthm}
\usepackage{color}
\def\identity{\leavevmode\hbox{\small1\kern-3.8pt\normalsize1}}
\newcommand{\beq}{\begin{equation}}
\newcommand{\eeq}{\end{equation}}
\newcommand{\beqa}{\begin{eqnarray}}
\newcommand{\eeqa}{\end{eqnarray}}

\begin{document}
\title{Off-Diagonal-Long-Range-Order Versus Entanglement}

\author{Dagomir Kaszlikowski}
\affiliation{Department of Physics, National University of
Singapore, 2 Science Drive 3, Singapore 117542}
\author{Marcin Wie\'sniak}
\affiliation{Department of Physics, National University of
Singapore, 2 Science Drive 3, Singapore 117542}

\date{\today}

\begin{abstract}
In this short note we discuss the relation between the so-called Off-Diagonal-Long-Range-Order in many-body interacting quantum systems introduced by C. N.
Yang in Rev. Mod. Phys. {\bf 34}, 694 (1962) and 
entanglement. We argue that there is a direct relation between these two concepts. 
\end{abstract}
\maketitle

We all know that entanglement is useful in quantum information processing. In principle, it is possible to built a
machine (quantum computer) that will outperform any classical computing device. 

Recently, we have started to realize that entanglement might also be related to thermodynamic properties of 
many-body systems \cite{fazio} although we are far from having a complete understanding of this relation.

In this short note we try to establish yet another evidence linking entanglement with thermodynamic properties of large 
many body systems in the context of the so-called Off-Diagonal-Long-Range-Order (ODLRO), the concept introduced in the 
Ref. \cite{yang}.  

The paper is organized as follows. In the first section we discuss the concept of entanglement of massive particles with vacuum. The aim
of this section is to show that this kind of entanglement is physical, i.e., it can be used for quantum information processing. 

The second section discusses a particular example of ODLRO, namely, the one existing in the BEC phase of a non-interacting bosonic gas.

Finally, in the third section we define in general the concept of ODLRO and argue that it is naturally linked to entanglement.  

\section{Entanglement of massive particles with vacuum}

Let us consider an infinite square well potential with a single particle in the first excited state inside the potential. In the second quantization formalism
this can be written as $a^{\dagger}_1|\Omega\rangle$, where the operator $a^{\dagger}_1$ creates the particle in the first excited state from the vacuum. If 
we now divide the well into half, creating a partition $A$ on the left and  a partition $B$ on the right from the centre of the well, we can formally write
\begin{equation}
a^{\dagger}_1|\Omega\rangle = \left(\frac{1}{\sqrt{2}}A_0^{\dagger} +\frac{1}{\sqrt{2}}B_0^{\dagger}\right)|\Omega\rangle,   
\end{equation}
where the operator $A_0^{\dagger}$ creates a particle in the $A$ partition described by the sinusoidal function being the part of the initial wave function and 
the operator  $B_0^{\dagger}$ creates a particle in the $B$ partition described by the other half of the initial wave function. 

The above state can be interpreted as a superposition of a particle being on the right (partition $B$) and no particle on the left (partition $A$) with the particle
being on the left (partition $A$) and no particle on the right (partition $B$), i.e.,

$$|1_A\rangle|\Omega_B\rangle+|\Omega_A\rangle|1_B\rangle.$$ The kets $|\Omega_A\rangle, |\Omega_B\rangle$ denote local vacuums. This kind of states spur 
lots of controversies regarding their entanglement status \cite{Hardy1,Hardy2,Hardy3,Hardy4}.
We do not want to discuss this issue in details here but we would like to mention that in the case of massless particles (photons for instance) there seems to be no
problem as the state can be used to entangle two distant atoms. In other words, for massless particles one can locally swap the entanglement in the state to two massive 
particles that later can be locally manipulated and used as a resource for quantum information processing. The photon is destroyed in the process.

It has been already shown in the Ref. \cite{marcello} that one can extract entanglement from a single massive particle as well and here we briefly present slightly expanded
and modified approach presented there. 

To this end imagine that we have two boxes each containing a single particle in the ground state of the box. The box $A$ will interact 
only with the partition $A$ and the box $B$ with the partition $B$. The interaction is given by  
\begin{equation}
V=(A_0C_AM_A^{\dagger}+A_0^{\dagger}C_A^{\dagger}M_A)+(B_0C_BM_B^{\dagger}+B_0^{\dagger}C_B^{\dagger}M_B),
\end{equation}
where the operators $C_A$ and $C_B$ destroy the particle in the box $A$ and $B$ and the operators $M_A^{\dagger},M_B^{\dagger}$ create a molecule in
the respective box. The molecule is created from the merger of the particle in the box and the particle in the well. The hamiltonian of the whole system consists of the free evolutions 
of each subsystem and the interaction $V$. We assume that the initial state is given by
\begin{equation}
|\psi_i\rangle = a_1^{\dagger}C_A^{\dagger}C_{B}^{\dagger}|\Omega\rangle,
\end{equation}
i.e., there is one particle in each box and there are no molecules.

In our scenario, we rapidly switch on and off the coupling (Dirac's delta coupling) $V$ such that the whole evolution reads
\begin{eqnarray}
&&|\psi_f\rangle=e^{i g V}|\psi_i\rangle =\nonumber\\
&& \left(\cos(g)A_0^{\dagger}C_A^{\dagger}C_B^{\dagger}+  \cos(g)B_0^{\dagger}C_A^{\dagger}C_B^{\dagger}+\right.\nonumber\\
&&\left. i\sin(g)C_B^{\dagger}M_A^{\dagger}+i\sin(g)C_A^{\dagger}M_B^{\dagger}\right)|\Omega\rangle.
\end{eqnarray}
In the Fock representation, the final state reads
\begin{eqnarray}
&&|\psi_f\rangle =\frac{1}{\sqrt{2}} \cos(g)\left(|1,0\rangle+|0,1\rangle\right)|1,0,1,0\rangle+\nonumber\\
&&\frac{1}{\sqrt{2}}i\sin(g)|0,0\rangle\left(|0,1,1,0\rangle+|1,0,1,0\rangle\right),
\end{eqnarray}
where the ket $|k,l,m,n,o,p\rangle$ denotes $k$ particles in the partition $A$, $l$ particles in the partition $B$, $m$ particles in the box $A$, $n$ molecules in the box $A$, $o$ particles in the box $B$ and $p$ molecules in the box B. Before tracing out the particle in the well to obtain the reduced density matrix of the system of particles and molecules  in the boxes $A$ and $B$ let us introduce a convenient notation. The ket representing one particle in the box $A$ ($B$) and no molecules in the box $A$ ($B$) is denoted by $|\uparrow\rangle_{A(B)}$ and the ket representing no particles in the box $A$ ($B$) and one molecule in the box $A$ ($B$) by $|\downarrow\rangle_{A(B)}$. In this notation, the reduced density matrix after the particle in the well has been traced out reads
\begin{eqnarray}
&&\cos(g)^2|\uparrow\rangle_A{}_A\langle\uparrow| \otimes|\uparrow\rangle_B{}_B\langle\uparrow|+\nonumber\\
&&\sin(g)^2|\psi_{-}\rangle_{AB}{}_{AB}\langle\psi_{-}|,
\end{eqnarray}
where $|\psi_{-}\rangle = \frac{1}{\sqrt{2}}\left(|\uparrow\rangle_A|\downarrow\rangle_B+|\downarrow\rangle_A|\uparrow\rangle_B\right)$. The negativity of this state is given by
$$\frac{1}{2}\left(\cos(g)^2-\sqrt{\cos(g)^4+\sin(g)^4}\right)$$ and is always non-zero as long as $g$ does not equal to the multiplicity of $\pi$. For $g=\frac{\pi}{2}$ we get maximal 
entanglement. The same extraction scheme works for an arbitrary wave function describing the particle in the box. 

Interestingly, after the interaction extracting the maximal amount of entanglement the particle in the well disappears in the exact analogy with the scheme 
of extracting entanglement from a single photon. 

We see that in this simple scenario we have locally extracted entanglement carried by a single massive particle and swapped it to a composite system of two local subsystems each containing massive particles of different kind. To verify this entanglement one must be able to measure the phase between the superposition of two different particles as mentioned in the Ref. \cite{marcello}. This can be done with the methods described in the Ref. \cite{rudolph}.

\section{BEC phase transition}
The presented scenario shows that the entanglement of a single massive particle is as physical as any other form of entanglement, which has been already argued in the Ref. \cite{marcello}. Let us know apply this results to a system of non-interacting spinless bosons in some three dimensional trapping potential. 

It is known, see the Ref. \cite{revzen} for instance, that the essential properties of a non-interacting gas of bosons near the point of the BEC phase transition are contained in the single particle density matrix. For a canonical ensemble of $N$ bosons, this density matrix reads
\begin{equation}
\rho_1 = \sum_{\vec{k}}\frac{\langle n_{\vec{k}}\rangle}{N} |\phi_{\vec{k}}\rangle\langle\phi_{\vec{k}}|,  
\end{equation}
where the vector $\vec{k}$ labels the eigenstates of the potential (in the case of the three dimensional box these are the momenta of the particle) and $\langle n_{\vec{k}}\rangle$ is the mean number of bosons occupying the energy level associated with $\vec{k}$. For each $\vec{k}$ the wave function
$ |\phi_{\vec{k}}\rangle$ can be written as 
\begin{equation}
|\phi_{\vec{k}}\rangle = \frac{1}{\sqrt{2}}|A_{\vec{k}}\rangle |\Omega_B\rangle+\frac{1}{\sqrt{2}}|\Omega_A\rangle |B_{\vec{k}}\rangle.
\end{equation}
where the ket $|A_{\vec{k}}\rangle$ ($\langle A_{\vec{k}}|A_{\vec{k}}\rangle=1$) represents the part of the wave function $|\phi_{\vec{k}}\rangle$ located in the half of the box (partition $A$) and the ket $|\Omega_B\rangle$ represents the vacuum in the other half of the box (partition $B$), and vice versa.    

Now, we are ready to perform partial transposition on $\rho_1$ with respect to, say, partition $B$ of the box. This partial transposition has a well defined meaning as it detects extractable entanglement. We get 
\begin{eqnarray}
&&\rho_1^{T_B} = \frac{1}{2}\sum_{\vec{k}}\left(|A_{\vec{k}}\rangle\langle A_{\vec{k}}|\otimes|\Omega_B\rangle\langle \Omega_B|\right.\nonumber\\
&&\left.+|\Omega_A\rangle\langle \Omega_A|\otimes |B_{\vec{k}}\rangle\langle B_{\vec{k}}|\right)+\nonumber\\
&&\frac{1}{2}\left(|\Omega_A,\Omega_B\rangle\langle\chi|+|\chi\rangle\langle \Omega_A,\Omega_B|\right),
\end{eqnarray}
where the ket $|\chi\rangle$ reads
\begin{equation}
|\chi\rangle = \sum_{\vec{k}}\frac{\langle n_{\vec{k}}\rangle}{N}|A_{\vec{k}}\rangle|B_{\vec{k}}\rangle
\end{equation}
and the norm squared of $|\chi\rangle$ is given by
\begin{equation}
\langle\chi|\chi\rangle = \sum_{\vec{k},\vec{l}} \frac{ \langle n_{\vec{k}} \rangle \langle n_{\vec{l}}\rangle}{N^2}\langle A_{\vec{k}}|A_{\vec{l}}\rangle\langle B_{\vec{k}}|B_{\vec{l}}\rangle,
\end{equation}
where we have assumed that we deal with real wave functions.

The only negative eigenvalue of $\rho_1^{T_{B}}$ is $\frac{1}{2}\sqrt{\langle\chi|\chi\rangle}$ and it can be plotted against the temperature once one knows the overlaps $\langle A_{\vec{k}}|A_{\vec{l}}\rangle,\langle B_{\vec{k}}|B_{\vec{l}}\rangle$. However, it is already clear that the plot will sharply change around the critical temperature for condensation because of the dominance of the term $\frac{\langle n_{0}\rangle}{N}$ near the critical temperature for $N\rightarrow\infty$. This observation establishes a simple connection between a single particle entanglement and Bose-Einstein condensation. In other words, one can quantify the BEC phase transition in terms of single-particle entanglement but this is a simple consequence of the fact that the single density matrix is enough to describe the properties of the whole system near the point of the phase transition.  

We would like to stress here that the spatial entanglement and its connection to BEC have been investigated in the Refs. \cite{spatial,libby1,libby2,libby3} in the context of entanglement of all the particles in the gas distributed amongst two partitions. The conclusions in the Ref. \cite{libby3} concerning a non-interacting gas of bosons are exactly the same.

The presented results do not depend on the division of the potential into two parts as long as the partitions do not overlap. However, if the partitions $A$ and $B$ are not adjacent, i.e., there is a partition $C$ between them, the reduced single-particle density matrix will contain a sector of vacuum in both partitions $A$ and $B$, which will decrease the amount of entanglement. More precisely, the negativity in such a situation reads
\begin{equation}
\frac{1}{2}(q-\sqrt{4\langle\delta|\delta\rangle+q^2}),  
\end{equation}
where $$|\delta\rangle = \sum_{\vec{k}}\frac{\langle n_{\vec{k}}\rangle}{N}\sqrt{p^{(A)}_{\vec{k}}p^{(B)}_{\vec{k}}}|A_{\vec{k}}\rangle |B_{\vec{k}}\rangle,$$ $p^{(M)}_{\vec{k}}$ being the probability of finding a particle with the wave vector $\vec{k}$ in the partition $M$ and $$q=\sum_{\vec{k}}\frac{\langle n_{\vec{k}}\rangle}{N}p^{(C)}_{\vec{k}}.$$ If the gap between the partition $A$ and $B$, i.e., the size of the partition $C$, is too big $q$ could be much larger than $\langle\delta|\delta\rangle$ reducing the amount of entanglement. Interestingly, in the absolute zero temperature the negativity reads $$\frac{1}{2}p^{(C)}_{0}\left(1-\sqrt{1+4 \frac{  p^{(A)}_{0} p^{(B)}_{0} }{  p_0^{(C)}p_0^{(C)}}} \right),$$ 
where $p_0^{(M)}$ is the probability of finding a particle in the ground state in the partition $M$. This expression is never zero as long as there is a chance of finding a particle in the partitions $A$ and $B$.

\section{Relation of ODLRO to entanglement}
Let us know generalize the above results to demonstrate a link between the concept of ODLRO and entanglement. Imagine a system of $N$ interacting particles (bosons or fermions). According to the Ref. \cite{yang} one can identify the different thermodynamical phases of the system by looking at the spectral properties of the reduced $n$-body density matrices $\rho_n$. For what follows next we set the normalization of the density operators $\rho_n$ to $Tr(\rho_n)=\frac{N!}{n!}$. 

The first reduced density matrix to investigate is $\rho_1$. The ODLRO in this matrix is defined as the existence of an eigenvalue of the order of $N$. This automatically means that the other eigenvalues are small and $\rho_1$ can be written as 
\begin{equation}
\rho_1 = \alpha N |\psi\rangle\psi|+\rho_1'
\end{equation}
where $\rho_1'$ is a small positive operator and $\alpha$ is some positive number. As we did before, we can quantify the single-particle entanglement by the negativity of the partially transposed matrix $\rho_1$ with respect to some partition of the trapping potential into two disjoined parts. We can expect, as in the case of the BEC discussed earlier, that the entanglement will sharply 
change around the point where the spectrum $\rho_1$ becomes concentrated around the dominating eigenvalue.   

The intuitive understanding of this behaviour of entanglement is the following one. As was shown in the Ref. \cite{yang} the existence of the eigenvalue of the order of $N$ implies that
\begin{equation} 
\langle\vec{r}|\rho_1|\vec{s}\rangle\rightarrow \frac{\alpha N}{V}   
\end{equation}
for $|\vec{r}-\vec{s}|\rightarrow\infty$, where $V$ is the volume occupied by the system of $N$ bosons. The above equation states that the amplitude of the probability of the process describing hopping of a particle
from the point $\vec{r}$ to the point $\vec{s}$ becomes substantial. This, in turn, implies that one can expect to have a single-particle entanglement between these two points in space;  one has the superposition of the particle at point $\vec{r}$ and no particle at point $\vec{s}$ with the particle at point $\vec{s}$ and no particle at point $\vec{r}$. Therefore, whenever there is ODLRO in the single-particle reduced density matrix there is a substantial amount of the single-particle entanglement.

One can also talk about ODLRO present in the two-body reduced density matrix without ODLRO in the one-body density matrix \cite{yang}. Its presence is detected by the existence of an eigenvalue of the order of $N$ with all the other ones being of the lower order (a good physical example of this kind is low temperature superconductivity, the BCS state).  In this case we can write 
\begin{equation}
\rho_2 = \alpha N |\phi\rangle\langle\phi| +\rho_2',
\end{equation}
where $|\phi\rangle$ is some two particle wave function and, as before, the $\rho_2'$ is a small positive operator. In the position representation this condition means \cite{yang} that only the diagonal elements of $\rho_2(\vec{r}_1\vec{r}_2|\vec{r}_1\vec{r}_2)$ and off diagonal ones of the form $\rho_2(\vec{r}\vec{r}|\vec{s}\vec{s})$ are non-zero, even for $|\vec{r}-\vec{s}|\rightarrow\infty$. Strictly speaking, this happens for $r$ and $s$ vectors microscopically close to each other; they do not have to point to the same point in space. Physically, this condition means that the amplitude of probability of two particles hopping from one location in space to another one dominates all the other processes. As the consequence, we expect to have entanglement generated between two arbitrary partitions. This entanglement will be of the form
\begin{equation}
|\phi\rangle_A|0\rangle_B+|0\rangle_A|\phi_B\rangle, 
\end{equation}
where the ket $|\phi_A\rangle$ represents the part of the two-particle ket $|\phi\rangle$ describing two particles in the partition $A$ (the same applies to the partition $B$). Additionally, this entanglement becomes substantial only in the regime where ODLRO in $\rho_2$ exists, for instance in the superconducting states.

In this short note we have presented a straightforward connection between the concept of ODLRO for a system of $N$ interacting bosons or fermions and entanglement. It seems that ODLRO implies entanglement of massive particles with vacuum (at least in the cases discussed in this note) but we do not know if the presence of  this kind of entanglement generates ODLRO. It would be interesting to investigate higher order ODLRO, i.e., the one present in $n$-body reduced density matrices and its relation to entanglement. This will be the subject of further research.

\acknowledgments
DK would like to thank Howard Wiseman, Dieter Jaksch and Berge Englert for interesting discussions. DK and MW are supported by the NUS grant R-144-000-206-112. MW acknowledges FNP Start stipend.

\end{document}